\documentclass{report}
\usepackage{svcon2e,graphicx,psfrag,natbib,amsmath}
\begin{document}
\pagenumbering{arabic}

\renewcommand\bibsection{\section*{R{\small EFERENCES}}}

\newcommand\dist{\buildrel\rm d\over\sim}
\newcommand\mis{_{\rm mis}}
\newcommand\com{_{\rm com}}
\renewcommand\c{_{\rm com}}
\newcommand\bin{_{\rm bin}}
\newcommand\obs{_{\rm obs}}
\newcommand\aug{_{\rm aug}}
\newcommand\Ya{\dot Y}
\newcommand\Yb{\ddot Y}
\newcommand\Yc{\dot{\ddot Y}\hskip -0.8pc\phantom{Y}}

\newcommand\lamc{\tilde{\cal F}}
\newcommand\alc{\alpha}
\newcommand\lam{\lambda}
\newcommand\itoN{^N_{i=1}}
\newcommand\iton{^n_{i=1}}
\newcommand\jtoJ{^J_{j=1}}
\renewcommand\th{\theta}
\newcommand\ind{\stackrel{\rm indep.}{\sim}}
\newcommand\updot{^{\textstyle\cdot}}
\newcommand\downdot{_{\textstyle\cdot}}
\newcommand\ds{\displaystyle}
\renewcommand\r{\right}
\renewcommand\l{\left}
\newcommand\Var{{\rm Var}}
\newcommand\E{{\rm E}}
\newcommand\cur{^{(t)}}
\newcommand\pre{^{(t-1)}}
\newcommand\nex{^{(t+1)}}

\newcommand{\btheta}{\boldsymbol{\theta}}
\newcommand{\bphi}{\mbox{\boldmath{$\phi$}}}
\newcommand{\bY}{\boldsymbol{Y}}
\newcommand\bX{\boldsymbol{X}} 

\def\Level{L{\scriptsize EVEL }}
\def\Step{S{\scriptsize TEP}}

\newcommand\spacingset[1]{\renewcommand{\baselinestretch}%
{#1}\small\normalsize}



\chapter{Bayesian Spectral Analysis of ``MAD'' Stars.}

\chapterauthors{Nondas Sourlas\footnotemark[1], David A. van Dyk
\footnote{Department of Statistics, Harvard University, One Oxford
St., Cambridge, MA 02138},
Vinay Kashyap\footnotemark[2], Jeremy Drake\footnotemark[2], and Deron Pease
\footnote{Harvard-Smithsonian
Center for Astrophysics, 60 Garden St., Cambridge, MA 02138}}.

\section{Overview}
Getting reliable estimates of coronal metallicity ($Z$) from X-ray
spectra obtained with instruments such as ASCA/SIS and Chandra/ACIS is
very difficult, because the sole determinant of $Z$ is the ratio of
line to continuum fluxes, which is not well-determined for
low-resolution spectra.  Here we propose new Bayesian methods which
directly model the Poisson nature of the data.  Our model also accounts
for the Poisson nature of background contamination, blurring due to 
instrument response, and the absorption of photons in space.  The 
resulting highly structured hierarchical model is fit using the Gibbs 
sampler, data augmentation, and Metropolis-Hasting.  We demonstrate our 
methods with the X-ray spectral analysis of several apparently coronal 
metal abundance deficient ("MAD") stars.

\section{A Poisson Spectral Model}
  
The model is designed to summarize the relative frequency of the
energy of photons arriving at a detector.  We model the photon counts 
in each bin as independent Poisson random variables.  Specifically, we 
model the high energy tail of the ASCA spectrum (2.5-7.5 keV) as a 
combination of a Bremsstrahlung continuum and ten narrow emission lines, 
included at positions of known strong lines.  This source model is 
combined with instrument response, the effective area of the instrument, 
and background contamination to model actual observed counts 
\citep{vand:conn:kash:siem:01}.

Statistical analysis is based on two observations of the same source
and two of the same background.  We use sequential Bayesian analysis
for the two source observations; the posterior distribution from the
first analysis is used to construct an informative prior for the
second.  Non-informative priors were used in the first stage.  Model 
fitting proceeds by using the EM algorithm to check for multimodality in 
the posterior.  A Markov chain Monte Carlo algorithm is constructed to 
sample from that posterior.  Three chains are used at each step of the 
analysis in order to assess convergence.  Posterior inference is based 
on the second half of the draws of all the chains.  The sensitivity of 
our results to the choice of prior was investigated by altering the prior.  
We used residual plots for both source and background to diagnose the fit 
of our models.

\section{Results}

We have measured the coronal metallicity for 4 stars, $\alpha$ Aur
(Capella), $\sigma$ Gem, YY Gem, and II Peg.  Our result for Capella 
($Z=0.73_{>0.24}^{<2.4}$) are consistent with the independently 
determined value of $Z=0.57-0.78$ \citep{brick:etal:00}.  Based on our 
analysis of $\sigma$ Gem ($Z=0.81_{>0.59}^{<1.1}$), we find that contrary 
to classical analyses ($Z=0.25$) it is {\sl not} a MAD star (photospheric 
$Z=0.6$, \citet{rand:etal:94}).  YY Gem is found to have sub-Solar 
metallicity ($Z=0.46_{>0.23}^{<0.71}$) but higher than EUVE/SW measurements 
($Z \sim 0.1$; \citet{kash:etal:98}).  For II Peg, we find $Z=1.1_{>0.88}^
{<1.3}$, higher by a factor $\sim 10$ than preliminary results from 
Chandra/HETG; this discrepancy may be due to anomalies in the abundances 
of other elements, or to a high-temperature component to the emission measure.

\vspace{0.5cm}

{\it Acknowledgments:}
The author gratefully acknowledges funding for this project partially
provided by NSF grant DMS-01-04129 and by NASA contract NAS8-39073
(CXC).  This chapter is a result of a joint effort of the members of the
Astro-Statistics working group at Harvard University.

\bibliography{../../../../TeXfiles/BibTeX/my}{

\begin{thebibliography}{}


\bibitem[van Dyk \emph{et~al.}(2001)van Dyk, Connors, Kashyap, and
  Siemiginowska]{vand:conn:kash:siem:01}
van Dyk, D.~A., Connors, A., Kashyap, V., and Siemiginowska, A. (2001).
\newblock Analysis of energy spectra with low photon counts via {B}ayesian
  posterior simulation.
\newblock \emph{The Astrophysical Journal} \textbf{548}, 224--243.

\bibitem[Brickhouse \emph{et~al.}(2000)Brickhouse, et al.]
{brick:etal:00}
Brickhouse, N.~S., Dupree, A.~K., Edgar, R.~J., Liedahl, D.~A.,
Drake, S.~A., White, N.~E., and Singh, K.~P. (2000)
\newblock \emph{The Astrophysical Journal} \textbf{530}, \textbf{387}.

\bibitem[Randich \emph{et~al.}(1994)Randich, Giampapa, and Pallavicini]
{rand:etal:94}
Randich, S., Giampapa, M.~S., and Pallavicini, R. (1994)
\newblock \emph{Astronomy and Astrophysics} \textbf{283}, \textbf{893}.

\bibitem[Kashyap \emph{et~al.}(1998)Kashyap, Drake, Pease and Schmitt]
{kash:etal:98}
Kashyap, V., Drake, J., Pease, D., and Schmitt (1998).
\newblock \emph{American Astronomical Society} \textbf{192}, 82.01.

\end{thebibliography}

\end{document}